\documentclass{emulateapj}

\usepackage{amssymb}
\usepackage{amsmath}
\usepackage{graphicx}
\usepackage{natbib}
\usepackage{txfonts}
\usepackage{microtype}
\usepackage{subfigure}
\usepackage{threeparttable}
\usepackage[colorlinks=true,citecolor=blue,linkcolor=blue]{hyperref}

\bibpunct{(}{)}{;}{a}{}{,}

\def\csic{1}
\def\ieec{2}
\def\upc{3}
\def\exeter{4}

\begin{document}

\title{A common origin of magnetism from planets to white dwarfs}

\date{Received 2016 October 20; Accepted 2017 February 4; Published 2017 February 17. Published in ApJL}

\author{Jordi~Isern\altaffilmark{\csic,\ieec},
        Enrique~Garc\'\i a--Berro\altaffilmark{\upc,\ieec},
        Baybars~K\"ulebi\altaffilmark{\csic,\ieec}, and
        Pablo Lor\'{e}n-Aguilar\altaffilmark{\exeter}}
\altaffiltext{\csic}{Institut de Ci\`encies de l'Espai (CSIC), 
                 Campus UAB,
                 08193 Cerdanyola, 
                 Spain}
\altaffiltext{\ieec}{Institut d'Estudis Espacials de Catalunya, 
                 Ed. Nexus-201, 
                 c/Gran Capit\`a 2-4, 
                 08034 Barcelona, 
                 Spain}
\altaffiltext{\upc}{Departament de F\'{i}sica, 
                 Universitat Polit\`{e}cnica de Catalunya, 
                 c/Esteve Terrades, 5, 
                 08860 Castelldefels, 
                 Spain}
\altaffiltext{\exeter}{School of Physics, 
                 University of Exeter, 
                 Stocker Road, 
                 Exeter EX4 4QL, 
                 UK}

\begin{abstract}
Isolated  magnetic  white dwarfs  have  field  strengths ranging  from
kilogauss to gigagauss. However, the  origin of the magnetic field has
not been hitherto elucidated.  Whether  these fields are fossil, hence
the remnants  of original  weak magnetic  fields amplified  during the
course of the  evolution of their progenitor stars, or  are the result
of  binary  interactions  or,  finally, they  are  produced  by  other
internal physical  mechanisms during  the cooling  of the  white dwarf
itself,  remains a  mystery.  At  sufficiently low  temperatures white
dwarfs crystallize. Upon solidification,  phase separation of its main
constituents, $^{12}$C  and $^{16}$O,  and of  the impurities  left by
previous evolution  occurs. This process  leads to the formation  of a
Rayleigh-Taylor unstable liquid  mantle on top of a  solid core.  This
convective region, as it occurs in Solar System planets like the Earth
and Jupiter,  can produce a  dynamo able  to yield magnetic  fields of
strengths  of up  to 0.1~MG,  thus  providing a  mechanism that  could
explain magnetism in single white dwarfs.
\end{abstract}


\section{Introduction}

White  dwarfs  are the  most  common  end-point of  stellar  evolution
\citep{alth10}.   Some of  them have  measurable magnetic  fields with
strengths  ranging  from  $10^3$   to  $10^9$~G  \citep{ferr15}.   The
incidence  of magnetism  is  a matter  of  debate.  In  volume-limited
surveys -- those that select stars  within a maximum distance from the
Sun  --  the incidence  of  magnetism  is $\sim  20\%$  \citep{kawk07,
giam12,  sion14},   while  for  magnitude-limited  surveys   --  those
selecting  stars brighter  than a  given apparent  magnitude --  it is
$\sim 8\%$  \citep{lieb03, kepl13,  kepl16}.  Nevertheless,  because a
population  of white  dwarfs with  magnetic fields  weaker than  $\sim
1$~kG may exist, the fraction of magnetic white dwarfs could be larger
\citep{kawk12,   koes11,    ferr15}.    Observations    also   suggest
\citep{valy99,  lieb03,  sion14,  kawk14,  valy14,  holl15}  that  the
fraction of white dwarfs with strong  magnetic fields is larger at low
effective temperatures.  This could  indicate that magnetic fields are
amplified during the evolution  of white dwarfs.  This interpretation,
however, has  been questioned \citep{ferr15} because  apparently there
is no clear correlation between the strength of the magnetic field and
the luminosity,  except for low  magnetic field ($B \lesssim  0.1$ MG)
white  dwarfs. Interestingly  enough,  the observed  paucity of  white
dwarfs with magnetic fields between 0.1 and 1~MG \citep{kawk12,koes11}
suggests a bimodal distribution \citep{ferr15}.

The origin  of magnetic white  dwarfs remains  unknown, and up  to now
three scenarios have  been proposed.  Within the first  of them, white
dwarf  magnetic fields  are  the  fossil remnants  of  those of  their
progenitors. Thus, the  progenitors of magnetic white  dwarfs could be
main  sequence Ap/Bp  stars \citep{ange81,  wick05}, or  could be  the
result  of   field  amplification  during  the   helium-burning  phase
\citep{levy74}.  Within the second scenario, magnetic white dwarfs are
the  result of  the evolution  of binary  systems.  In  this case  the
magnetic field  is amplified by  a dynamo that operates  either during
the common envelope phase \citep{tout08,  nord11} or in the hot corona
produced  during the  merger of  two white  dwarfs \citep{garc12}.   A
difficulty  of  these  two   scenarios  is  that  detailed  population
synthesis studies cannot reproduce the number of stars observationally
found \citep{ferr15}.  Finally, within the third scenario the field is
generated in the  outer convective envelope that is  formed during the
evolution of  single white dwarfs.  However,  theoretical calculations
show that  the strength of  the resulting field is  $\lesssim 0.01$~MG
\citep{font73,trem15}, well  below the observed magnetic  fields, so a
more efficient mechanism should be invoked.

The  evolution  of   white  dwarfs  can  be  described   as  a  simple
gravothermal process \citep{alth10}. The core  of the vast majority of
white dwarfs is  made of a completely ionized mixture  of $^{12}$C and
$^{16}$O,  and   some  minor  chemical  species   like  $^{22}$Ne  and
$^{56}$Fe. During  their evolution, two physical  processes modify the
internal  chemical  profiles  of  white  dwarfs.   The  first  one  is
gravitational  settling of  neutron-rich species  in the  liquid phase
\citep{brav92,bild01,garc10,cami16}, which  occurs at  moderately high
luminosities.  The second one is phase separation upon crystallization
\citep{iser97,iser00,garc10},    and    takes     place    at    lower
luminosities.  In both  cases,  the energy  involved  is large,  $\sim
2\times 10^{46}$~erg.

For the  sake of simplicity, here  we will only consider  white dwarfs
made of $^{12}$C and $^{16}$O.  The phase diagram of the carbon-oxygen
mixture is of the spindle  form \citep{segr93,horo10}. Since the solid
phase is oxygen-rich,  hence denser than the liquid  phase, when white
dwarfs crystallize it settles down,  while the carbon-rich liquid left
behind    is    redistributed   by    Rayleigh-Taylor    instabilities
\citep{iser97,iser00}.  This configuration, a solid core surrounded by
a convective  mantle driven  by compositional  buoyancy is  similar to
that found in  the core of the Earth, where  the light element release
associated  with the  inner core  growth is  a primary  driver of  the
dynamo \citep{list95}.  Given this analogy,  in this Letter we analyze
if the  convective mantle  of crystallizing  white dwarfs  can produce
magnetic fields of the observed strengths.

\section{The stellar dynamo}

For  a   typical  white  dwarf   of  mass  $\simeq   0.61\,  M_{\sun}$
solidification starts at $\log(L/L_{\sun}) \simeq -3.5$, whereas for a
massive one  of $1.0 \,  M_{\sun}$ crystallization sets in  at $\simeq
-2.6$. At this point of the evolution the temperatures of their nearly
isothermal cores are, respectively, $\log T_{\rm c}\, ({\rm K}) = 6.3$
and $6.6$,  whereas their central  densities are $\log  \rho_{\rm c}\,
({\rm g\, cm^{-3}})=  6.6$ and $7.5$ \citep{sala10}.  For  the sake of
conciseness, and unless specified otherwise, here we will only discuss
the case of the heavier white dwarf, since it is observationally found
that magnetic white dwarfs are more massive than usual \citep{ferr15}.

\begin{figure}
\plotone{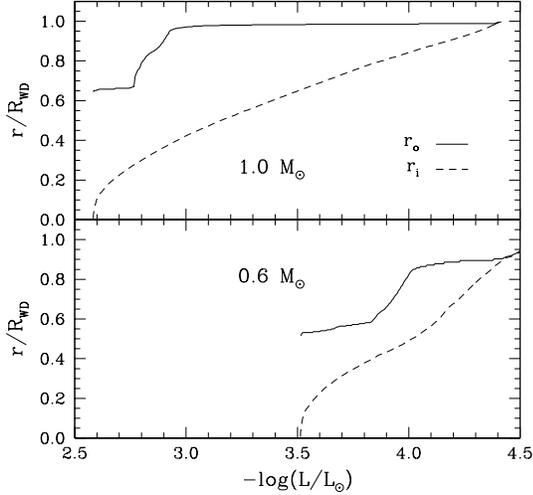}
\caption{Evolution  of the  inner and  outer radii  of the  convective
  mantle  of  a  carbon-oxygen  white  dwarf  as  a  function  of  the
  luminosity. The  upper and lower  panels correspond to  white dwarfs
  with  masses  $1.0$  and   $0.6\,  M_{\sun}$,  respectively.   Their
  respective total  radii are $R  = 4.7  \times 10^8$ and  $7.5 \times
  10^8$~cm.
\label{f1}}
\end{figure}

As can be  seen in Fig.~\ref{f1}, for a $1.0  \,M_{\sun}$ white dwarf,
when  solidification   starts  the  Rayleigh-Taylor   unstable  region
encompasses a  large fraction  of the  radius of  the star.  This also
holds for  the $0.6 \,M_{\sun}$  model star.  As the  $1.0 \,M_{\sun}$
white  dwarf cools,  the outer  edge of  the convective  mantle barely
moves,  while  the  inner   edge  moves  progressively  outwards.   At
$\log(L/L_{\sun})\simeq -2.9$, the outer edge of the mantle is located
at $r/R  \simeq 0.65$, and  the inner edge  at $\simeq 0.4$.   At this
moment,   the   outer   edge   moves  abruptly   outwards   and   when
$\log(L/L_{\sun})\simeq -3.1$ it almost reaches the surface, while the
inner  edge continues  its progression  at  a slower  pace.  The  main
difference with the $0.6 \,M_{\sun}$ white  dwarf is that in this case
the  outer radius  stabilizes at  $r/R\simeq 0.9$.   Finally, in  both
cases,  the convective  region disappears  at about  $\log(L/L_{\sun})
\simeq -4.5$.

The  density contrast  between the  carbon enriched  material and  the
ambient liquid mixture is $\delta \rho /\rho \sim 10^{-3}$, leading to
effective  accelerations $a_{\rm  eff} =  g(\delta \rho/\rho)\simeq  2
\times  10^6$~cm~s$^{-2}$. However,  since  the  viscosity of  Coulomb
plasmas  is very  small, drag  cannot be  neglected and  results in  a
limiting    velocity     of    the    turbulent     eddies,    $v_{\rm
s}=\sqrt{(3/8)C_{\rm b} a_{\rm eff} D_{\rm  b}}$, where $D_{\rm b}$ is
the radius of the bubble, and $C_{\rm b} \approx 0.2$ in the spherical
case \citep{garc95}.  The  value of $D_{\rm b}$ is not  yet known, but
it  cannot be  larger than  the curvature  radius at  the edge  of the
crystallized core.  When the same value found for the Earth is adopted
$D_{\rm b} \sim 0.1 R_{\rm core}$ \citep{moff94}, we obtain velocities
of $\sim 35$~km~s$^{-1}$.

For  the thermodynamical  conditions found  in white  dwarf interiors,
viscosities   are   very   small  and   conductivities   rather   high
\citep{nand84}.   In  particular, for  the  heavier  white dwarf,  the
electrical conductivity is $\sigma  = 1.3\times 10^{21}$~s$^{-1}$, the
magnetic diffusivity is  $\eta=5.6 \times 10^{-2}$~cm$^2$~s$^{-1}$ and
the kinematic viscosity is $\nu =3.13 \times 10^{-2}$~cm$^2$~s$^{-1}$.
Adopting $\sim  2\times 10^8$~cm  for the  characteristic size  of the
convective mantle  (see Fig.~\ref{f1}, upper panel),  the Reynolds and
magnetic Reynolds numbers\footnote{The  magnetic Reynolds number, Rm$=
ul/\eta$, where  $u$ and  $l$ are typical  velocities and  lengths and
$\eta$ is  the magnetic diffusivity, measures  the relative importance
of induction versus dissipation, while  the magnetic Prandtl number is
the  ratio between  the magnetic  and the  ordinary Reynolds  numbers,
Pm=Rm/Re.}  range from  $\sim 10^{14}$ to $10^{15}$,  and the magnetic
Prandtl  number is  $\approx 0.58$.   Furthermore, the  characteristic
Ohmic decay time $\tau_\Omega  \sim R^2_{\rm WD}/\eta\simeq 10^{18}$~s
is much  longer than the age  of the white dwarf.   Consequently, once
the magnetic  field is generated,  it will remain almost  constant for
long times \citep{cumm02}.

\begin{figure}
\plotone{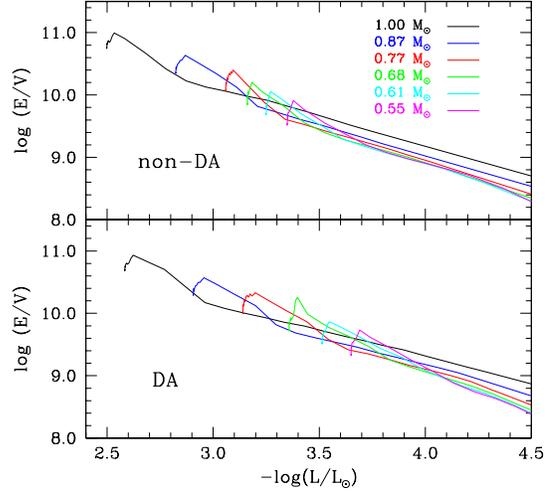}
\caption{Convective  energy  density available  for  the  dynamo as  a
  function of the  luminosity, for white dwarfs of  masses 1.00, 0.87,
  0.77, 0.68, 0.61,  and $0.55\, M_{\sun}$. The  top panel corresponds
  to  white dwarfs  with H-deficient  atmospheres (non-DA),  while the
  bottom one corresponds to stars with H-rich atmospheres (DA).
  \label{f2}}
\end{figure}

\begin{table*}[t]
\centering
\caption{Properties of magnetic white dwarfs within 20~pc of the Sun.}
\label{tb1}
\begin{threeparttable}
\begin{small}
\begin{tabular}{lccccccccc}
\hline
\hline
Name          &  Type  & $B$~(MG)           & Period~(h)           & $M\,(M_{\sun})$ & $\log(L/L_{\sun})$ & $M_{\rm s}/M_{\rm WD}$  & $\rho_{\rm b}$   & $r_{o}-r_{i}$      & $q_{\rm top}$       \\ 
\hline 
WD 0009+501   &     DA &         0.2        &  2 -- 20             & $0.73 \pm 0.04$ & $-3.72$            & 0.43                    & $3.0\times 10^6$ & $2.3\times 10^8$   & $7.1\times 10^{10}$ \\
WD 0011$-$134 &     DA & $16.7\pm 0.6$      &    ---               & $0.72 \pm 0.07$ & $-3.85$            & 0.52                    & $2.3\times 10^6$ & $2.6\times 10^8$   & $5.7\times 10^{10}$ \\
WD 0322$-$019 &     DA & 0.120              &    ---               & $0.63 \pm 0.05$ & $-4.02$            & 0.52                    & $1.6\times 10^6$ & $5.2\times 10^8$   & $3.0\times 10^{10}$ \\
WD 0413$-$077 &     DA & $0.0023\pm 0.0007$ &    ---               & $0.59 \pm 0.03$ & $-1.85$            & 0.00                    & ---              & ---                & ---                 \\
WD 0548$-$001 & non-DA & $\sim 10$          &   4.117              & $0.69 \pm 0.03$ & $-3.80$            & 0.82                    & $7.0\times 10^5$ & $1.2\times 10^8$   & $6.5\times 10^{10}$ \\
WD 0553+053   &     DA & $20\pm 3$          &   0.97               & $0.72 \pm 0.03$ & $-3.91$            & 0.35                    & $1.9\times 10^6$ & $2.5\times 10^8$   & $4.6\times 10^{10}$ \\
WD 0728+642   &     DA & $0.0396\pm 0.0116$ &    ---               & $0.58 \pm 0.00$ & $-4.00$            & 0.36                    & $1.5\times 10^6$ & $2.0\times 10^8$   & $3.7\times 10^{10}$ \\
WD 0912+536   & non-DA & $\sim 100$         &  31.9224             & $0.75 \pm 0.02$ & $-3.57$            & 0.71                    & $1.5\times 10^6$ & $1.6\times 10^8$   & $1.1\times 10^{11}$ \\
WD 1008+290   & non-DA & $\sim 100$         &    ---               & $0.68 \pm 0.01$ & $-4.31$            & 0.93                    & $9.8\times 10^4$ & $3.0\times 10^6$   & $2.1\times 10^{10}$ \\
WD 1036$-$204 & non-DA & 50                 & $636 \pm 36$         & $0.60 \pm 0.01$ & $-4.19$            & 0.76                    & $1.9\times 10^5$ & $1.1\times 10^7$   & $2.5\times 10^{10}$ \\
WD 1309+853   &     DA & $4.9\pm0.5$        &    ---               & $0.71 \pm 0.02$ & $-4.01$            & 0.60                    & $1.6\times 10^6$ & $2.3\times 10^8$   & $3.4\times 10^{10}$ \\
WD 1748+708   & non-DA & $\ga 100$          & $\ga 1.7\times 10^4$ & $0.79 \pm 0.01$ & $-4.07$            & 0.79                    & $2.0\times 10^5$ & $4.5\times 10^6$   & $4.6\times 10^{10}$ \\
WD 1829+547   & non-DA & $170-180$          &    ---               & $0.90 \pm 0.07$ & $-3.94$            & 0.85                    & $2.7\times 10^5$ & $7.2\times 10^6$   & $8.3\times 10^{10}$ \\
WD 1900+705   & non-DA & $320\pm20$  & $\ga 8.5 \times 10^5$       & $0.93 \pm 0.02$ & $-2.88$            & 0.18                    & $1.4\times 10^7$ & $2.4\times 10^8$   & $1.2\times 10^{12}$ \\
WD 1953$-$011 &     DA & $0.1-0.5$          & 34.60224             & $0.73 \pm 0.03$ & $-3.38$            & 0.13                    & $5.6\times 10^6$ & $2.4\times 10^8$   & $3.3\times 10^{11}$ \\
WD 2105$-$820 &     DA & $0.010\pm 0.001$   &    ---               & $0.74 \pm 0.13$ & $-2.93$            & 0.00                    & ---              & ---                & ---                 \\
WD 2359$-$434 &     DA & $0.0031\pm0.0005$  &    ---               & $0.78 \pm 0.03$ & $-3.26$            & 0.12                    & $7.2\times 10^6$ & $1.8\times 10^8$   & $7.3\times 10^{11}$ \\
\hline
\end{tabular}
\end{small}
\begin{tablenotes}[para,flushleft]
Data obtained from \citet{kawk07}, \citet{giam12} and references therein.
\end{tablenotes}
\end{threeparttable}
\end{table*}

The aim  of dynamo  scaling theories  is to  explain the  intensity of
magnetic fields in  terms of the properties of the  region hosting the
dynamo.  The most  comprehensive of these theories  takes into account
the  balance  between  the  ohmic  dissipation  and  the  energy  flux
available to  the dynamo  \citep{chri10}. The  magnetic fields  of the
Earth and Jupiter  are generated by convective dynamos  powered by the
cooling and  chemical segregation of  their interiors and it  has been
shown that the scaling law based  on the energy fluxes can be extended
to fully convective  stars like T~Tauri and  rapidly rotating M~dwarfs
\citep{chri09}.  Therefore,  it is  natural to  use these  theories to
compute  the  magnetic field  produced  by  the convective  mantle  of
crystallizing white dwarfs.

If  the dynamo  is saturated,  the magnetic  Reynolds number  is large
enough, and  convection is described  by the mixing  length formalism,
this scaling law can be written as \citep{chri10}:
\begin{equation}
\frac{{B^2 }}{{2\mu _0 }} = cf_\Omega  
\frac{1}{V}\int\limits_{r_i }^{r_o } 
{\left[ {\frac{{q_c \left( r \right)\lambda\left( r \right)}}
{{H\left( r \right)}}} \right]}^{{2\mathord{\left/{\vphantom {2 3}} \right. \kern- \nulldelimiterspace} 3}}
\,\rho \left( r \right)^{{1 \mathord{\left/
 {\vphantom {1 3}} \right.
 \kern-\nulldelimiterspace} 3}} 4\pi r^2 dr
\label{chris_rel}
\end{equation}
where the integral is the energy of the convective mantle, $E$, $c$ is
an adjustable parameter, $\mu_0$ is the vacuum permeability, $f_\Omega
\le 1$ is the ratio of the Ohmic dissipation to the total dissipation,
$q_{\rm c}$  is the  convected energy  flux, $H$  is the  scale height
(temperature and compositional),  $V$ is the volume  of the convective
region, $r_o$ and  $r_i$ are its outer and inner  radii, and $\lambda$
is the mixing length (the minimum between the density scale height and
the  size  of the  convective  zone,  $r_o  -  r_i$).  Here  we  adopt
$f_\Omega  = 1$  and  $\lambda=H$.   Also, we  use  the BasTI  cooling
sequences \citep{sala10} which,  in addition to the  release of latent
heat,  consider  the  energy   released  by  chemical  differentiation
\citep{moch83,segr94}.

\section{Results and discussion}

\begin{figure}
\plotone{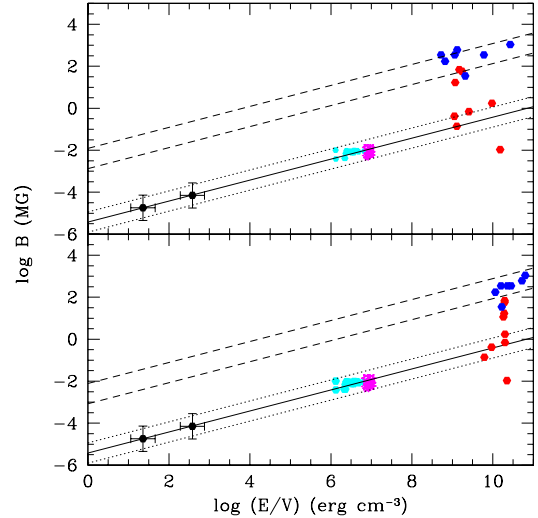}
\caption{Magnetic field intensity  as a function of  the dynamo energy
  density. Earth and Jupiter are  represented using black symbols with
  their corresponding error bars. T~Tauri and M~dwarfs are shown using
  cyan and  magenta symbols, respectively \citep{chri09}.   The DA and
  non-DA white dwarfs listed in Table~\ref{tb1} are plotted as red and
  blue  symbols, respectively.   The  solid line  is the  relationship
  relating  the magnetic  fields of  the Earth,  Jupiter, T~Tauri  and
  M~dwarf stars, while dotted lines  allow for an additional deviation
  of a factor of  3 from it. The dashed lines  help to represent where
  non-DA  stars  cluster.  The  top  and  the bottom  panels  display,
  respectively, the intensity  of the magnetic field as  a function of
  the present and of the maximum energy densities of the dynamo.
\label{f3}}
\end{figure}

Figure~\ref{f2} displays the density of convective energy available to
the dynamo, namely the  r.h.s.  of Eq.~(\ref{chris_rel}) excluding the
factor $cf_\Omega$, for several white dwarf masses.  The dynamo starts
when  solidification sets  in, then  reaches a  maximum and,  finally,
slowly declines as the bottom  of the convective mantle moves outwards
at fainter luminosities.   This is the result of  two effects.  First,
the mass  of the  star, that  determines when  crystallization starts.
White  dwarfs with  larger masses  crystallize at  larger luminosities
because their  central densities are  also larger.  The  luminosity of
the star  also plays  a role,  since it determines  the rate  at which
crystallization takes place.  This luminosity, in turn, depends on the
transparency of the  atmosphere and the temperature of  the core.  The
maximum energy available  to the dynamo ranges from $\log  E = 8.2$ to
9.6,  and from  8.3 to  9.7 (c.g.s.   units) for  DA and  non-DA white
dwarfs,   respectively,    when   masses   ranging   from    0.54   to
$1.00\  M_{\sun}$  are  considered.   The  scaling  law  relating  the
magnetic fields of  the Earth, Jupiter, T~Tauri and  M~dwarf stars can
be approximated by $\log B = 0.5  \log E - 5.42$ -- see Fig.~\ref{f3},
solid  line  -- and  predicts  that  the  maximum  field that  can  be
generated at the top of the dynamo range from $\sim 0.05$ to 0.25~MG.

The X-ray  emission of stars  with convective regions  correlates with
the rotation period,  and with magnetic activity.  The  existence of a
plateau  in  the  relationship  linking magnetic  activity  and  X-ray
luminosity  suggests  that dynamos  saturate  when  the Rossby  number
Ro$=P_{\rm rot}/\tau<0.1$,  being $\tau$ the convective  turnover time
\citep{wrig11,wrig16}.   For a  star of  mass $1.0  \, M_{\sun}$,  and
adopting the  typical values  found in Sect.~2,  the turnover  time is
$\sim 90$~s.   Since the critical  stability rotation period  is $\sim
6$~s,  we  conclude  that  rapidly  rotating  white  dwarfs  can  have
saturated  dynamos.  Because  white  dwarf spectral  lines are  broad,
measuring the rotation periods is a difficult task, and have only been
measured  in  pulsating  and   magnetic  white  dwarfs.   Thus,  their
distribution   of  angular   velocities   is   not  well   determined.
Observations  indicate  that  white  dwarfs  are  slow  rotators,  but
strictly  speaking, the  existence  of rapidly  rotating white  dwarfs
cannot be discarded \citep{kawa15}.   Within our scenario white dwarfs
with  high magnetic  fields would  rotate rapidly  and have  saturated
dynamos, while  stars with  weak fields would  rotate slowly  and have
non-saturated dynamos.  Therefore, our  mechanism could easily explain
white dwarfs with magnetic fields $B \lesssim 0.1$~MG.

Table~\ref{tb1}   shows   the   properties  of   white   dwarfs   with
carbon-oxygen cores within 20~pc  of the Sun \citep{kawk07,giam12}. We
list the name of the star,  the spectral type, the magnetic field, the
rotation period in hours, the mass  and luminosity in solar units, the
relative  mass of  the crystallized  core  and the  properties of  the
convective mantle  (the density at the  inner edge, the size,  and the
flux at  the top, all  in c.g.s. units).   It is important  to realize
that all  of them,  except WD~0413$-$077 (40~Eri~B)  and WD~2105$-$820
(L24-52) have crystallized cores.  For each of these stars we computed
the   energy   density   available   to  the   dynamo   according   to
Eq.~(\ref{chris_rel})  employing the  value of  $c$ derived  using the
scaling law for the Earth and Jupiter, T~Tauri stars and M~dwarfs.  In
the top  panel of  Fig.~\ref{f3} we  show the  set of  calculations in
which the energy density of the  dynamo has been computed adopting the
actual size  of the  convective mantle  corresponding to  the observed
luminosity  of the  star,  whereas in  the bottom  panel  we show  the
magnetic field strength when the  maximum energy density of the dynamo
shown in  Fig.~\ref{f2} is  adopted. This  is a  reasonable assumption
given the very  long ohmic decay timescale.  In both  cases, to relate
the  outer and  inner  magnetic  fields we  used  the prescription  of
\citet{chri09}.   The  distribution  of magnetic  field  strengths  is
clearly bimodal. Most  DA white dwarfs cluster around  the scaling law
(the  solid line  in Fig.~\ref{f3}),  while non-DA  white dwarfs  have
magnetic fields  larger than DA  stars with similar energy  density of
the dynamos.  This points towards to a different origin of the dynamo,
and/or to  other scaling  laws like  the Coriolis-Inertial-Archimedian
balance or  the Magnetic-Archimedian-Coriolis  balance \citep{chri10},
because the  mixing length scaling  law considered here  demands $\sim
10^{15}$~erg~s$^{-1}$ to produce fields of $\sim100$~MG.

It is  important to realize that  the time necessary for  the magnetic
field to diffuse through the radiative  outer layers is very long, and
only the  convective mantle of  massive white  dwarfs is close  to the
surface, as  clearly shown in  Fig~\ref{f1}. The final  magnetic field
strength will depend on how  this mantle interacts with the convective
envelope of  cool white  dwarfs. Furthermore,  only white  dwarfs with
progenitors  more  massive than  $2.3\,  M_{\sun}$  can have  rotation
periods of minutes \citep{kawa15}.  This,  together with the fact that
the  energy of  the dynamo  induced by  crystallization is  larger for
massive white dwarfs could naturally explain why magnetic white dwarfs
are more massive  than average.  This scenario could  also explain why
magnetic white dwarfs are preferentially cool.  Moreover, it is likely
that low-mass white dwarfs would  have weaker magnetic fields confined
in  their   interiors.   Therefore,  this  scenario   puts  forward  a
tantalizing possibility, namely, that  the magnetic fields observed in
planets, non-evolved  stars and white  dwarfs share a  common physical
origin.  In summary, our calculations indicate that the magnetic field
observed in a sizable fraction of all white dwarfs could be the result
of a dynamo generated by phase separation upon crystallization.

Furthermore, our  scenario does not preclude  other possibilities, but
instead alleviates one of the main drawbacks of the current hypotheses
to explain  magnetic white dwarfs,  since both the fossil  field model
and  the binary  scenario predict  an insuficient  number of  magnetic
white dwarfs.  For instance, it has  been long suspected that at least
a fraction  of high-field magnetic  white dwarfs could  originate from
the merger of  two white dwarfs \citep{wick00}.  During  the merger, a
hot,   differentially-rotating,   convective    corona   forms.    The
temperatures reached during the coalescence  are so high that hydrogen
is burned during the early phases of the merger.  This corona is prone
to  magnetorotational instability,  and  it has  been  shown that  can
produce  magnetic  fields with  the  energy  required to  explain  the
magnetic fields of non-DA stars shown in Fig.~\ref{f3} \citep{garc12}.
This  could explain  why many  white  dwarfs with  very high  magnetic
fields  are  H-deficient.  In  this  case  a  dynamo of  a  completely
different nature would be operating.

\acknowledgements
This work has been supported by MINECO grants ESP2013- 47637-P, ESP2015-66134-R (J.I.), and AYA2014-59084-P (E.G.-B.), by the European Union FEDER funds, by grants 2014SGR1458 (J.I.), 2014SGR0038 (E.G.-B.) of the AGAUR, and by the CERCS program of the Generalitat de Catalunya.


\begin{thebibliography}{}
\bibitem[Althaus  et  al.(2010)]{alth10}  Althaus, L.  G.,  C\'orsico,
  A. H., Isern, J., \& Garc\'{\i}a-Berro, E. 2010, A\&A Rev., 18, 471
\bibitem[Angel et al.(1981)]{ange81} Angel, J.~R.~P., Borra, E.~F.,
  \& Landstreet, J.~D.\ 1981, \apjs, 45, 457
\bibitem[Bildsten  \&  Hall(2001)]{bild01}   Bildsten,  L.,  \&  Hall,
  D.~M.\ 2001, \apjl, 549, L219
\bibitem[Bravo et al.(1992)]{brav92} Bravo,  E., Isern, J., Canal, R.,
  \& Labay, J.  1992, \aap, 257, 534
\bibitem[Camisassa et  al.(2016)]{cami16} Camisassa, M.   E., Althaus,
  L. G., C\'orsico, A. H., et al. 2016, \apj, 823, 158
\bibitem[Christensen   et  al.(2009)]{chri09}   Christensen,  U.   R.,
  Holzwarth, V., \& Reiners, A. 2009, Nature, 457, 167
\bibitem[Christensen  (2010)]{chri10} Christensen,  U. R.  2010, Space
  Sci. Rev., 152, 565
\bibitem[Cumming(2002)]{cumm02} Cumming, A. 2002, \mnras, 333,589
\bibitem[Ferrario et al.(2015)]{ferr15} Ferrario,  L., de Martino, D.,
  \& G\"ansicke, B. T. 2015, Space Sci. Rev., 191, 111
\bibitem[Fontaine et al.(1973)]{font73} Fontaine, G., Thomas, J. H. \&
  Van Horn, H. M. 1973, \apj, 184, 911
\bibitem[Garcia-Berro  et  al.(2010)]{garc10}  Garc\'{\i}a-Berro,  E.,
  Torres, S., Altahus, L.G. et al. 2010 Nature, 465, 194
\bibitem[Garcia-Berro    et   al.(2012)]{garc12}    Garc\'{\i}a-Berro,
  Lor\'en-Aguilar., L.,  {Aznar-Sigu{\'a}n}, G.   et al.   2012, \apj,
  749, 25
\bibitem[Garc\'{\i}a-Senz \&  Woosley(1995)]{garc95} Garc\'{\i}a-Senz,
  D. \& Woosley, S.E. 1995, \apj, 454, 895
\bibitem[Giammichele  et al.(2012)]{giam12}Giammichele,  N., Bergeron,
  P., Dufour, P. 2012 \apjs, 199, 29
\bibitem[Hollands  et al.(2015)]{holl15}  Hollands,  M. A.,  Gansicke,
  B. T., Koester, D. 2015, \mnras, 450, 681
\bibitem[Horowitz  et al.(2010)]{horo10}  Horowitz, C.  J., Schneider,
  A.S., \& Berry, D. K. 2010, \prl, 104, 231101
\bibitem[Isern  et  al.(1997)]{iser97}  Isern,  J.,  Mochkovitch,  R.,
  Garc\'{\i}a-Berro, E., \& Hernanz, M. 1997, \apj, 485, 308
\bibitem[Isern et al.(2000)]{iser00} Isern, J., Garc\'{\i}a-Berro, E.,
  Hernanz, M., \& Chabrier, G. 2000, \apj, 528, 397
\bibitem[Kawaler(2015)]{kawa15}  Kawaler,  S.  D. 2015,  19th  EUROWD,
  ASPC, 493,65
\bibitem[Kawka et  al.(2007)]{kawk07} Kawka, A., Vennes,  S., Schmidt,
  G. O., et al. 2007, \apj, 654, 499
\bibitem[Kawka \& Vennes(2012)]{kawk12} Kawka,  A. \& Vennes, S. 2012,
  \aap, 538,13
\bibitem[Kawka \& Vennes(2014)]{kawk14} Kawka,  A. \& Vennes, S. 2014,
  \mnras, 439, L90
\bibitem[Kepler  et al.(2013)]{kepl13}  Kepler, S.  O., Pelisoli,  I.,
  Jordan, S. et al. 2013, \mnras, 429, 2934
\bibitem[Kepler  et  al.(2016)]{kepl16}  Kepler, S.O.,  Pelisoli,  I.,
  Koester, D., et al. 2016, \mnras, 455, 3413
\bibitem[Koester  et  al.(2011)]{koes11}   Koester,  D.,  Girven,  J.,
  G\"ansicke et al. 2011, \aap, 530, 114
\bibitem[Levy \&  Rose(1974)]{levy74} Levy,  E.H. \& Rose,  W.K. 1974,
  \apj, 193, 419
\bibitem[Liebert  et  al.(2003)]{lieb03}  Liebert, J.,  Bergeron,  P.,
  Holberg, J. 2003, \aj, 125,348
\bibitem[Lister \& Buffet(1995)]{list95} Lister, J.R. \& Buffet, B.A. 1995, PEPI, 91,17
\bibitem[Moffatt \& Loper(1994)]{moff94} Moffatt, H.K. \& Loper, D.E. 1994, Geophys. J. Int., 117, 394
\bibitem[Mochkovitch(1983)]{moch83} Mochkovitch,  R. 1983,  \aap, 122,
  212
\bibitem[Nandkumar   \&  Pethick(1984)]{nand84}   Nandkumar,  R.,   \&
  Pethick, C.J. 1984, \mnras, 209, 511
\bibitem[Nordhaus  et  al.(2011)]{nord11}  Nordhus, J.,  Wellons,  S.,
  Spiegel,   D.S.,   Metzger,   B.D.   \&  Blackman,   E.   G.   2011,
  Proc. Natl. Acad. Sci., 108, 3135
\bibitem[Salaris  et  al.(2010)]{sala10}  Salaris,  M.,  Cassisi,  S.,
  Pietrinferni, A., Kowalski, P.M., \& Isern, J. 2010, \apj, 716, 1241
\bibitem[Segretain  \&   Chabrier(1993)]{segr93}  Segretain,   L.,  \&
  Chabrier, G. 1993, \aap, 271, L13
\bibitem[Segretain et al.(1994)]{segr94}  Segretain, L., Chabrier, G.,
  Hernanz, M. et al. 1994, \apj, 434, 641
\bibitem[Sion et al.(2014)]{sion14} Sion, E.M., Holberg, J.B., Oswalt,
  T.D., McCook, G.P. Wasatonic, R., Myska, J. 2014, \aj,147, 129
\bibitem[Tout  et  al.(2008)]{tout08}  Tout,  C.  A.,  Wickramasinghe,
  D. T.,  Liebert, J.,  Ferrario, L., \&  Pringle, J.E.  2008, \mnras,
  387, 897
\bibitem[Tremblay et al.(2015)]{trem15} Tremblay, P.~E., Fontaine, G.,
  Freytag, B., Steiner, O., Ludwig, H.~G. et al. 2015, \apj, 812, 19
\bibitem[Valyavin \&  Fabrika(1999)]{valy99} Valyavin, G.  \& Fabrika,
  S. 1999, Astron. Soc. Pacif. Conf. Ser., 169, 206
\bibitem[Valyavin  et al.(2014)]{valy14}  Valyavin,  G., Shulyak,  D.,
  Wade, G. A. et al. 2014, Nature, 515, 2934
\bibitem[Wickramasinghe  \&   Ferrario(2000)]{wick00}  Wickramasinghe,
  D. T., \& Ferrario, L. 2000, \pasp, 112, 873
\bibitem[Wickramasinghe  \&   Ferrario(2005)]{wick05}  Wickramasinghe,
  D. T., \& Ferrario, L. 2005, \mnras, 356, 1576
\bibitem[Wright  \&   Drake(2011)]{wrig16}  Wright,  N.J.   \&  Drake,
  J. J. 2016, Nature, 535, 526
\bibitem[Wright  et  al.(2011)]{wrig11}  Wright,  N.J.,  Drake,  J.J.,
  Mamajek, E.E. \& Henry, G.H. 2011, \apj, 743, 48

\end{thebibliography}
\end{document}